\documentclass{PoS}

\usepackage{graphicx}
\usepackage{subfigure}
\usepackage{amssymb}
\usepackage{amsmath}
\usepackage{mathrsfs}
\usepackage{dsfont}
\usepackage{amsfonts}

\title{Leading-order hadronic contributions to $a_\mathrm{\mu}$ and $\alpha_\mathrm{QED}$ from $N_f=2+1+1$ twisted mass fermions}

\ShortTitle{$a_\mu^\mathrm{hvp}$ and $\Delta \alpha_\mathrm{QED}^\mathrm{hvp}$ from $N_f=2+1+1$ twisted mass fermions}

\author{Xu Feng\\
        High Energy Accelerator Research Organization (KEK), Tsukuba 305-0801, Japan\\
        E-mail: \email{xufeng@post.kek.jp}}

\author{\speaker{Grit Hotzel} \\
        Humboldt-Universit\"at zu Berlin, Institut f\"ur Physik, D-12489 Berlin, Germany\\
        E-mail: \email{grit.hotzel@physik.hu-berlin.de}}

\author{Karl Jansen\\
        NIC, DESY, Plantanenallee 6, D-15738 Zeuthen, Germany\\
        E-mail: \email{karl.jansen@desy.de}}

\author{Marcus Petschlies\\
        The Cyprus Institute, P.O. Box 27456, 1645 Nicosia, Cyprus\\
        E-mail: \email{m.petschlies@cyi.ac.cy}}

\author{Dru B. Renner\\
        Jefferson Lab, 12000 Jefferson Avenue, Newport News, VA 23606, USA\\
        E-mail: \email{dru@jlab.org}}

\abstract{We present the first four-flavour lattice calculation of the leading-order 
hadronic vacuum-polarisation contribution to the anomalous magnetic moment of the
muon, $a_\mathrm{\mu}^{\rm hvp}$, and the hadronic running of the QED coupling constant,  $\Delta \alpha_\mathrm{QED}^\mathrm{hvp} (Q^2)$. In the heavy sector a mixed-action setup is employed. 
The bare quark masses are determined from matching the K-
and D-meson masses to their physical values. Several light quark masses are used in order to yield a controlled extrapolation to the physical pion
mass by utilising a recently proposed improved method. 
We demonstrate that this method also works in the four-flavour case.}

\FullConference{The 30th International Symposium on Lattice Field Theory\\
		 June 24 - 29,  2012\\
		 Cairns, Australia}

\begin{document}

\section{Introduction}
\label{sec:introduction}
The anomalous magnetic moment of the muon, 
$a_{\mathrm{\mu}}$,  serves as a benchmark test of the standard model (SM).
It has been  
measured very accurately~\cite{Bennett:2006fi,Roberts:2010cj} and 
can be computed precisely within the SM.
 A comparison between the experimental 
result for $a_{\mathrm{\mu}}$ and the SM 
prediction reveals a discrepancy of more than 
three standard deviations 
which has persisted for many years
now and has been confirmed by computations of a number of groups, 
see, for example, the review~\cite{Jegerlehner:2009ry}. 
The question is whether this discrepancy originates
from some effect missing in the experimental or theoretical
determination of $a_{\mathrm{\mu}}$ or
whether it points 
to physics beyond the SM.

A key ingredient in the calculation 
of $a_\mathrm{\mu}$ is 
the leading-order hadronic vacuum-polarisation contribution, 
$a_\mathrm{\mu}^{\rm hvp}$, which presently is the largest source
of uncertainty in the theoretical computation of
$a_\mathrm{\mu}$, since the QED and electroweak contributions have been computed very accurately employing perturbation theory, see~\cite{Aoyama:2012wk, Jegerlehner:2009ry} and references therein. As $a_\mathrm{\mu}^{\rm hvp}$ is intrinsically 
nonperturbative, a lattice QCD computation of this 
observable is highly desirable. The currently accepted SM values for this 
quantity are obtained mainly from the investigation of $e^+ e^-$ 
scattering and $\tau$ decay data. 
In a recent study~\cite{Feng:2011zk,Renner:2012fa}, using 
two flavours of mass-degenerate quarks, a modified method to compute $a_\mathrm{\mu}^{\rm hvp}$ has been introduced resulting in a determination of $a_{\mathrm{\mu}, \mathrm{light}}^{\rm hvp}$ with a precision of a few percent.

Besides $a_\mathrm{\mu}^{\rm hvp}$, the leading-order QCD contribution 
to the running of the QED coupling constant, $\Delta \alpha_\mathrm{QED}^\mathrm{hvp}$, 
also requires the hadronic vacuum-polarisation function 
and can likewise be investigated once this function
is known. As an important input parameter to SM calculations, the QED 
coupling constant needs to be known very precisely in order to facilitate high precision tests of the SM at any future linear collider~\cite{Jegerlehner:2011mw}.

Below we report preliminary results of the first lattice calculation of $a_\mathrm{\mu}^{\rm hvp}$ and $\Delta \alpha_\mathrm{QED}^{\rm hvp}$
with four quark flavours.
The contributions of the top and bottom quarks are negligible at the current level of accuracy. So having four flavours allows for an unambiguous
comparison to the dispersive analysis of $a_\mathrm{\mu}^{\rm hvp}$ and direct use in forming the SM prediction for $a_\mathrm{\mu}$ itself.
The inclusion of the charm quark is essential because its contribution is of the order of $a_\mathrm{\mu}^\mathrm{hvp, charm} \gtrsim 100 \times10^{-11}$~\cite{Jegerlehner:2011ti}, which is
larger than the currently quoted uncertainty of the difference between the experimental and the SM results. Furthermore, the order of magnitude of the charm quark contribution to $a_\mathrm{\mu}^{\rm hvp}$ is the same as
that of the hadronic light-by-light contribution~\cite{Prades:2009tw}. Thus lattice calculations with a dynamical charm quark are
necessary for computing $a_\mathrm{\mu}^{\rm hvp}$ with a precision that matches the experimental accuracy.

The computation of $a_\mathrm{\mu}^{\rm hvp}$ and $\Delta \alpha_\mathrm{QED}^{\rm hvp}$ follows closely 
the strategy of refs.~\cite{Feng:2011zk, Renner:2012fa} using improved lattice definitions 
of these quantities. We demonstrate in this work that this new method continues to work well even for the four-flavour
case and again results in a mild quark mass dependence leading to an accurate determination
of $a_\mathrm{\mu}^{\rm hvp}$ and $\Delta \alpha_\mathrm{QED}^{\rm hvp}$.

Our calculations 
are based on the configurations with four dynamical quark flavours generated by the 
European Twisted Mass Collaboration (ETMC) \cite{Baron:2010bv,Baron:2010th}. 
These sets 
of configurations are obtained at different values of the lattice spacing 
and several lattice volumes, 
thus enabling us to estimate discretisation and finite size
effects 
as systematic uncertainties of our lattice calculation. 
In addition, at each value of the lattice spacing, configurations exist 
at several values of the pion mass, ranging from 
$230\,\mathrm{MeV} \lesssim m_\pi \lesssim 450\,\mathrm{MeV}$.

\section{The lattice calculation}
\label{sec:numerics}

The leading-order hadronic contribution to the muon's anomalous magnetic 
moment, $a_{\mathrm{\mu}}^{\mathrm{hvp}}$, 
can be computed directly in Euclidean space-time~\cite{Blum:2002ii}
\begin{equation}
a_{\mathrm{\mu}}^{\mathrm{hvp}} = 
\alpha^2 \int_0^{\infty} \frac{d Q^2 }{Q^2} 
w\left( \frac{Q^2}{m_{\mathrm{\mu}}^2}\right) \Pi_{\mathrm{R}}(Q^2)\; \mathrm{.}
\label{eq:hvpdef}
\end{equation}
 The renormalised vacuum-polarisation function is given by
$\Pi_{\mathrm{R}}(Q^2)= \Pi(Q^2)- \Pi(0)$. 
$\Pi(Q^2)$ can be obtained from the 
hadronic vacuum-polarisation tensor knowing its Lorentz structure
\begin{equation}
   \Pi_{\mu \nu}(Q)= \int d^4 x \,e^{iQ\cdot(x-y)} \langle J_{\mu}^{\mathrm{em}}(x) J_{\nu}^{\mathrm{em}}(y)\rangle = (Q_{\mu} Q_{\nu} -Q^2 \delta_{\mu
\nu} ) \Pi(Q^2) \; \mathrm{,}
\label{eq:vptensor}
\end{equation}
where 
\begin{equation}
J_{\mu}^{\mathrm{em}}(x) = 
\frac{2}{3} \overline{u}(x) \gamma_{\mu} u(x) - \frac{1}{3} \overline{d}(x) \gamma_{\mu} d(x) + \frac{2}{3}
\overline{c}(x) \gamma_{\mu} c(x) - \frac{1}{3} \overline{s}(x) \gamma_{\mu} s(x) 
\label{eq:current}
\end{equation}
is the electromagnetic vector current.

On the lattice we
use the conserved (point-split) vector current
\begin{eqnarray}
 J_{\mu}^C(x)  & = & \frac{1}{2} 
                     \left( \overline{\chi}(x+\hat{\mu})(\mathds{1}
                     +\gamma_{\mu}) U_{\mu}^{\dagger}(x)Q_\mathrm{el} \chi(x) \right. \nonumber \\
 &  - & \overline{\chi}(x)(\mathds{1}-\gamma_{\mu}) U_{\mu}(x)Q_\mathrm{el} \chi (x+\hat{\mu}) \Big) \; \mathrm{.}
\label{eq:conscurrent}
\end{eqnarray}
$\chi(x)$ denotes a fermion doublet, either the light or the heavy one, in the so-called twisted basis and $Q_\mathrm{el} =
\mathrm{diag}(\frac{2}{3}, -\frac{1}{3})$ denotes the electric charge matrix. 
Resorting to Osterwalder-Seiler valence quarks~\cite{Osterwalder:1977pc,Frezzotti:2004wz} in the heavy sector admits
a straightforward construction of the conserved currents also for the strange and the charm quark. This is beneficial since in this way we can rely on the vector Ward-Takahashi identity for all contributions and
also avoid renormalisation.

In order to have a smooth function to perform the integral in eq.~(\ref{eq:hvpdef}), 
the vacuum-polarisation data 
obtained at discrete lattice momenta is fit for each flavour to the following functional form
\begin{equation}
  \Pi(Q^2) = g_{\rm V}^2 \frac{m_{\rm V}^2}{Q^2+m_{\rm V}^2} + b_0 + b_1 Q^2\; .
\label{eq:fitform}
 \end{equation}
The first term is the contribution of a narrow-width vector meson with mass $m_{\rm V}$ and electromagnetic coupling $g_{\rm V}$, which are determined directly in
our calculation. The remaining terms parametrise any deviations from this form.
The results reported below are obtained using uncorrelated fits to determine
$g_{\rm V}$ and $m_{\rm V}$ from the zero-momentum current correlators.
In these proceedings we do not perform alternative 
fits such as the Pad\'e approximants suggested in~\cite{Aubin:2012me}. 
We also do not provide an estimate of systematic effects by allowing 
additional terms in eq.~(\ref{eq:fitform}).  
These aspects will be addressed in a forthcoming publication. 

Once $\Pi_{\mathrm{R}}(Q^2)$ is known, it is straightforward to compute the
leading-order hadronic contribution~\cite{Jegerlehner:2011mw}
\begin{equation}
\Delta \alpha_\mathrm{QED}^{\mathrm{hvp}}(Q^2) = 4 \pi \alpha_0 \Pi_{\mathrm{R}}\left(Q^2\right)
\end{equation}
which influences the running of the fine structure constant according to~\cite{Jegerlehner:1985gq}
\begin{equation}
 \alpha_\mathrm{QED}(Q^2) = \frac{\alpha_0}{1-\Delta\alpha_\mathrm{QED}(Q^2) }\; \mathrm{.}
\end{equation}
Here, $\alpha_0$ is the value at $Q^2=0$, $\alpha_0 \approx \frac{1}{137}$.

For the lattice calculation of $a_{\mathrm{\mu}}^{\mathrm{hvp}}$ discussed here, we use the modified definition from~\cite{Feng:2011zk}
\begin{equation}
 a_{\overline{\mathrm{\mu}}}^{\mathrm{hvp}} = \alpha_0^2 \int_0^{\infty} \frac{d Q^2 }{Q^2} w\left( \frac{Q^2}{H^2}
\frac{H_{\mathrm{phys}}^2}{m_{\mathrm{\mu}}^2}\right) \Pi_{\mathrm{R}}(Q^2) \; \mathrm{,}
\label{redef}
\end{equation}
which goes to 
$a_{\mathrm{\mu}}^{\mathrm{hvp}}$ for the light pseudoscalar mass 
$m_{PS}$ assuming its physical value, i.e.~$m_{\pi}$, since in this case also $H \rightarrow H_{\mathrm{phys}}$, 
with the choice of the hadronic scale, $H$, discussed below.  Analogously, we determine $\Delta \alpha_\mathrm{QED}^{\mathrm{hvp}}(Q^2)$ from~\cite{Renner:2012fa}
\begin{equation}
 \Delta \overline{\alpha}_{\mathrm{QED}}^{\mathrm{hvp}}(Q^2) = 4 \pi \alpha_0 \Pi_{\mathrm{R}}\left(Q^2 \frac{H^2}{H_{\mathrm{phys}}^2}\right) \mathrm{.}
 \label{redefalpha}
\end{equation}

There are several possible choices for $H$. Below we give results 
for $H=H_{\rm phys}\equiv 1$ (the standard choice) and 
$H=m_{\rm V}(m_{\rm PS})$, $H_{\rm phys}=m_\rho$ (improved choice). 
Here $m_{\rm V}(m_{\rm PS})$ is the light vector meson mass as measured
on the lattice at unphysical pseudoscalar masses $m_{\rm PS}$ whereas 
$m_\rho$ is the physical value of the $\rho$-meson mass. 
See refs.~\cite{Feng:2011zk,Renner:2012fa} for a more detailed discussion.

\begin{figure}[ht]
\centering
\includegraphics[width=0.5\textwidth]{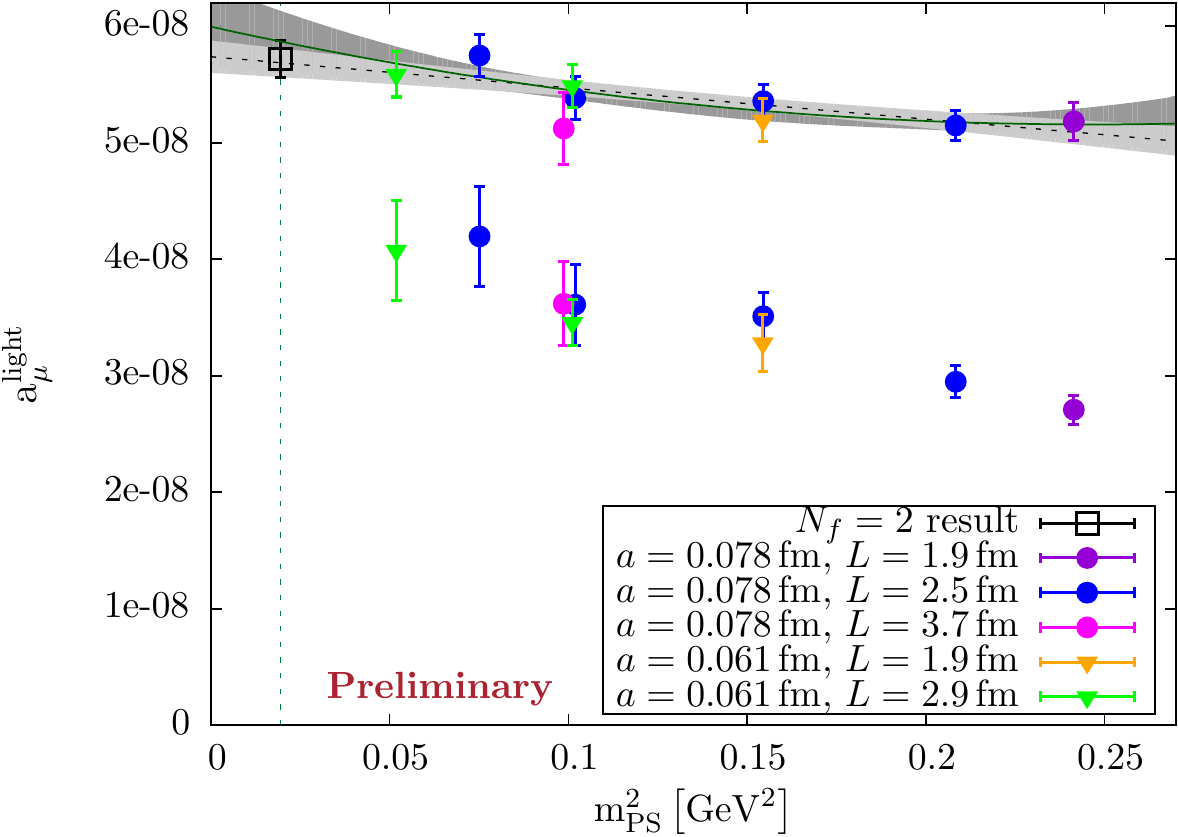}
\caption{Light quark contribution to 
 $a_{\mathrm{\mu}}^{\rm hvp}$ as a function of 
the squared light pseudoscalar mass from our $N_f=2+1+1$ computations. 
The lower set of data points
 corresponds to using the standard definition, 
 $H=H_{\rm phys}=1$ in eq.~(\protect\ref{redef}). The upper data points are obtained using the improved method, setting 
 $H=m_{\rm V}(m_{\rm PS})$ and $H_{\rm phys} = m_\rho$. 
 The open square represents the two-flavour result of refs.~\cite{Feng:2011zk, Renner:2012fa}. The dark grey error band corresponds to the quadratic fit (solid green line) and the light grey one belongs to the linear fit (dashed black line).}
\label{fig:amulight}
\end{figure}
 
The light quark contribution to 
$a_{\mathrm{\mu}}^{\mathrm{hvp}}$ is depicted in fig.~\ref{fig:amulight}. In contrast to the standard choice the 
improved lattice definition of $a_{\mathrm{\mu}}^{\mathrm{hvp}}$ shows 
a weaker pion mass dependence. Comparing the previous calculation 
with only two dynamical flavours of light, mass-degenerate quarks~\cite{Feng:2011zk} to the one
having four flavours in the sea quark sector, 
the results for $a_{\mathrm{\mu},\mathrm{light}}^{\mathrm{hvp}}$ extrapolated
to the physical point are found to be in full agreement 
for both setups. 
This demonstrates that the effects of a dynamical strange and charm quark 
on the light quark contribution to $a_{\mathrm{\mu}}^{\mathrm{hvp}}$ are 
small, as expected.

\begin{figure}[ht]
\centering
\includegraphics[width=0.5\textwidth]{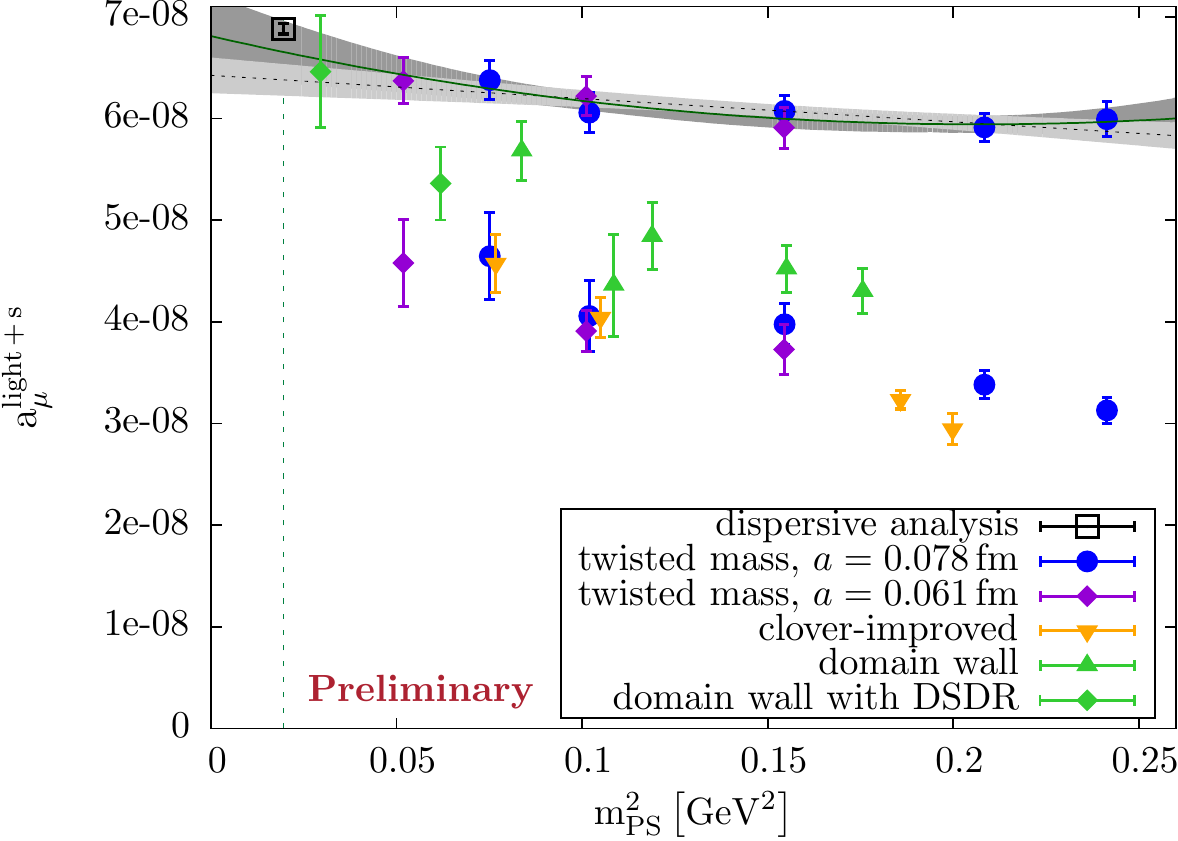}
\caption{Comparison of the results for 
$a_{\mathrm{\mu}}^{\mathrm{hvp}}$ having the light up and 
down quarks as well as the strange quark in the valence sector for different 
collaborations. Only the upper set of data points for twisted mass fermions are obtained from the modified definition,  
$H=m_{\rm V}(m_{\rm PS})$ and $H_{\rm phys} = m_\rho$
in eq.~(\protect\ref{redef}). Data are from~\cite{DellaMorte:2011aa} (clover-improved) and~\cite{Boyle:2011hu} (domain wall). 
The open square represents the standard model value 
obtained from the dispersive analysis of ref.~\cite{Benayoun:2011mm}. Concerning the error bands of the fits, the same comments as in fig.~\protect\ref{fig:amulight} apply.} 
\label{fig:amu2+1}
\end{figure}

In fig.~\ref{fig:amu2+1} we show the effect of adding the strange 
quark contribution and compare to the published results of other groups,
\cite{Boyle:2011hu, DellaMorte:2011aa}.
The figure demonstrates that when using the standard definition
of $a_{\mathrm{\mu}}^{\mathrm{hvp}}$ on the lattice, all groups 
agree reasonably well. However, when using the improved definition, 
the pion mass dependence of $a_{\mathrm{\mu}}^{\mathrm{hvp}}$
is much flatter allowing for a better control of the extrapolation 
to the physical pion mass. 

\begin{figure}[ht]
\centering
\includegraphics[width=0.5\textwidth]{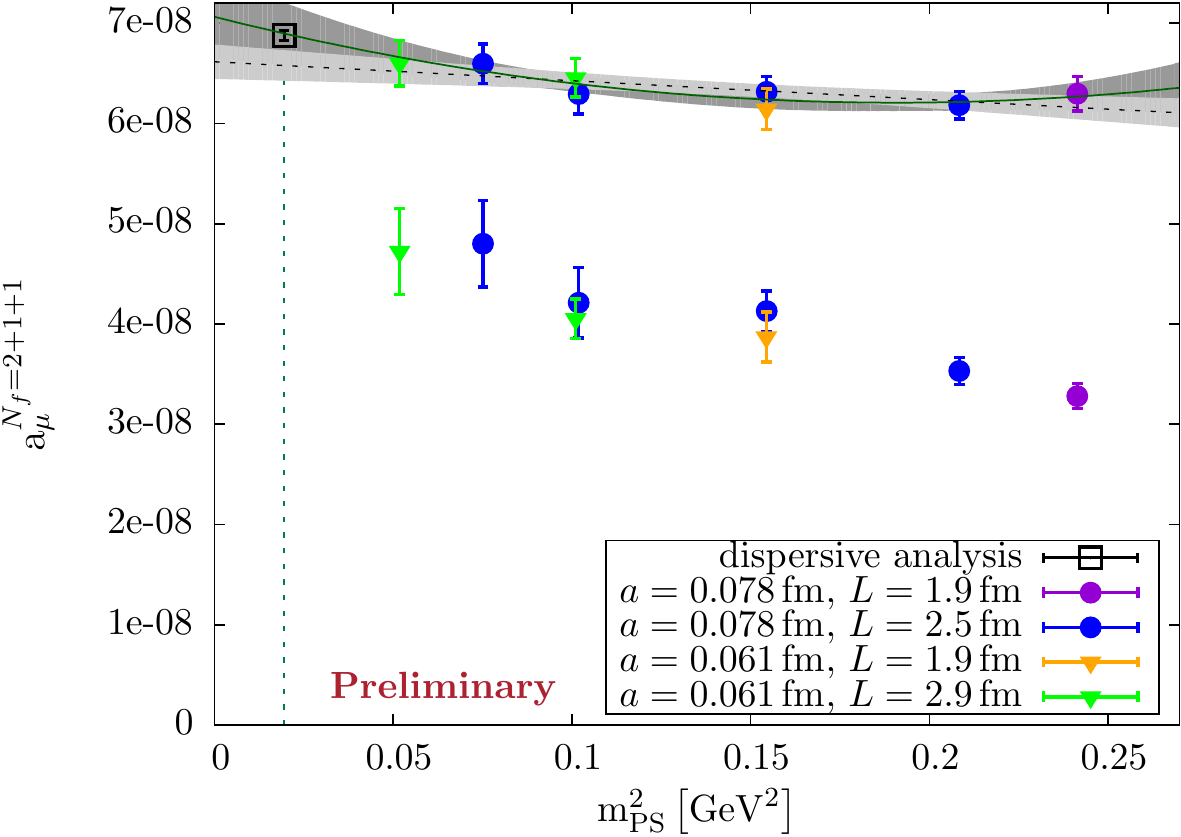}
\caption{Full four-flavour contribution to 
$a_{\mathrm{\mu}}^{\mathrm{hvp}}$ as a function of 
the squared light pseudoscalar mass from our $N_f=2+1+1$ computations.
Concerning the upper and lower data sets and the error bands of the fits, the same comments as in fig.~\protect\ref{fig:amulight} apply.
The open square represents the standard model value 
obtained from the dispersive analysis of ref.~\cite{Benayoun:2011mm}.
Note that in this case of four flavours there is no 
ambiguity in determining the contribution
to $a_{\mathrm{\mu}}^{\mathrm{hvp}}$ from the dispersive analysis.} 
\label{fig:amu2+1+1}
\end{figure}

Finally, we show in fig.~\ref{fig:amu2+1+1} the full four-flavour 
contribution to $a_{\mathrm{\mu}}^{\mathrm{hvp}}$. 
Comparing with fig.~\ref{fig:amu2+1}, we see that including the charm quark indeed 
leads to a contribution of the expected order of magnitude.
We perform both a linear and a quadratic extrapolation to the physical point. Within just the statistical errors, we
find reasonable agreement with the dispersive result~\cite{Benayoun:2011mm}, as shown in fig.~\ref{fig:amu2+1+1}. At the current precision, we do not observe any
statistically significant finite-size or lattice discretisation effects, however, the impact of disconnected contributions or
unitary-violating effects due to the mixed-action setup in the heavy sector have not been examined yet. Furthermore, the effect from the $\rho$-meson
not being able to decay to two pions for all but one ensemble has to be investigated. The details of the final extrapolation to the physical point and
the associated systematic uncertainties will be presented elsewhere.

\begin{figure}[ht]
\centering
\includegraphics[width=0.5\textwidth]{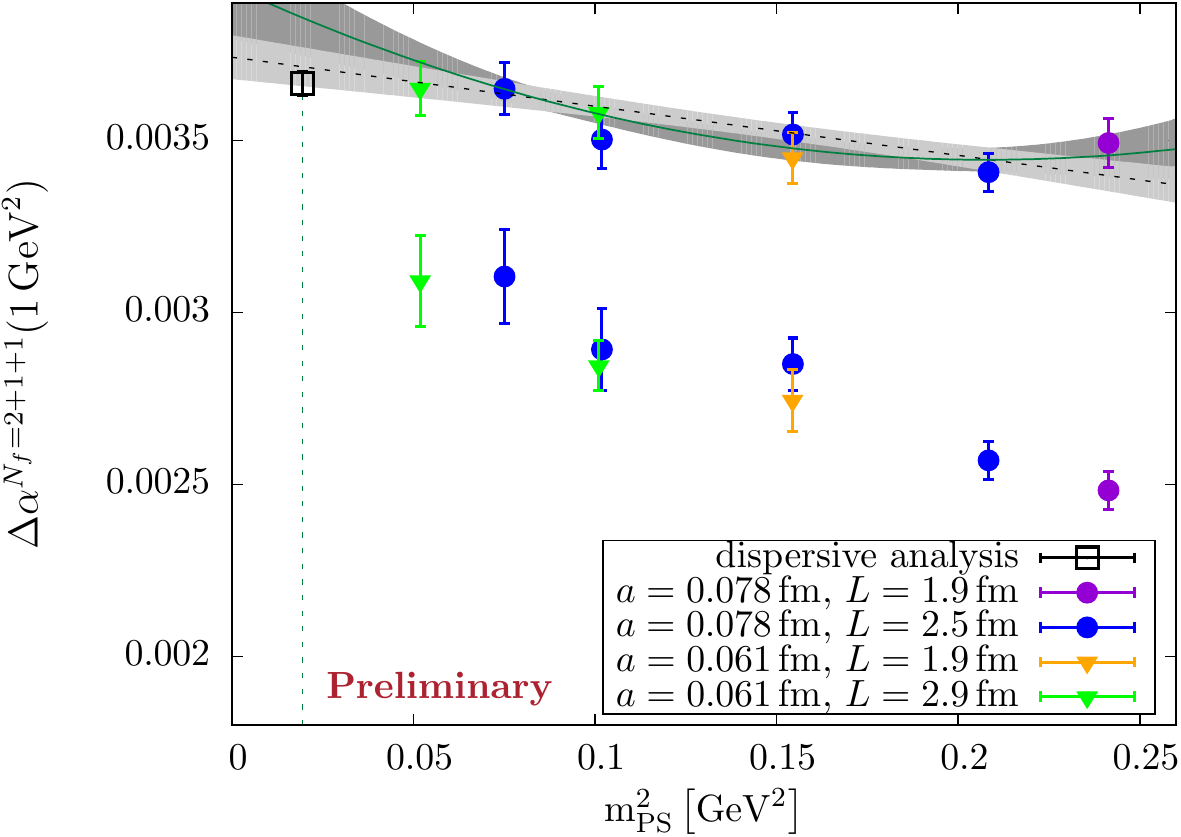}
\caption{Full four-flavour contribution to $\Delta \alpha_\mathrm{QED}^{\rm hvp}(1\, \mathrm{GeV}^2)$ as a function of 
the squared light pseudoscalar mass from our $N_f=2+1+1$ computations. 
The lower set of data points
corresponds to the use of the standard definition, 
$H=H_{\rm phys}=1$ in eq.~(\protect\ref{redefalpha}). The upper data points are obtained using the improved method, setting 
$H=m_{\rm V}(m_{\rm PS})$ and $H_{\rm phys} = m_\rho$. Concerning the error bands of the fits, the same
comments as in fig.~\protect\ref{fig:amulight} apply.
The open square represents the standard model value 
obtained from the dispersive analysis of ref.~\cite{Jegerlehner:2011mw}.
Note that in this case of four flavours there is no 
ambiguity in determining the contribution
to $\Delta \alpha_\mathrm{QED}^{\rm hvp}(1\, \mathrm{GeV}^2)$ from the dispersive analysis.} 
\label{fig:alpha2+1+1}
\end{figure}

Fig.~\ref{fig:alpha2+1+1} shows our results for $\Delta \alpha_\mathrm{QED}^{\rm hvp}$ at a typical hadronic scale of $Q^2=1\, \mathrm{GeV}^2$
obtained from the same ensembles used to determine $a_{\mathrm{\mu}}^{\mathrm{hvp}}$. Similar to $a_{\mathrm{\mu}}^{\mathrm{hvp}}$, we find reasonable
agreement with the dispersive result~\cite{Jegerlehner:2011mw} to within the statistical uncertainties. A full determination of the systematic
uncertainty for the extrapolated value will also be presented in a later publication.

\section{Conclusion}

We have pointed out that the charm quark contribution is necessary for achieving a lattice computation of $a_{\mathrm{\mu}}^{\mathrm{hvp}}$ with a precision that is comparable to the experimental one. Furthermore, we have shown that also in the case of a four-flavour calculation 
of the leading-order hadronic contribution  to
the muon anomalous magnetic moment and to the running of the electromagnetic coupling constant, the improved method 
of ref.~\cite{Feng:2011zk, Renner:2012fa} continues to work well.
We have so far not performed a comprehensive 
analysis of the systematic uncertainties.
In particular, the disconnected 
contributions originating from the strange quark might be non-negligible. Additionally, it might be necessary to take isospin breaking effects into account as the precision of our computation improves.

We conclude by emphasising that the strategy followed here can also 
be applied to other observables such as the Adler function, the corrections to the energy levels of muonic and ordinary hydrogen, and the weak mixing angle, as discussed in~\cite{Renner:2012fa}.

\section*{Acknowledgements}
We thank Andreas Nube for performing the matching of the K- and D-meson masses in the mixed-action setup with their physical values.
This work has been supported in part by the DFG Corroborative
Research Center SFB/TR9.
G.H.~gratefully acknowledges the support of the German Academic National Foundation (Studienstiftung des deutschen Volkes e.V.) and of the
DFG-funded Graduate School GK 1504.
This manuscript has been coauthored by Jefferson Science Associates, LLC under Contract No.~DE-AC05-06OR23177 with the U.S.~Department of Energy.
The numerical computations have been performed on the
{\it SGI system HLRN-II} at the {HLRN Supercomputing Service Berlin-Hannover},  FZJ/GCS
and BG/P at FZ-J\"ulich.

\end{document}